# Pattern formation in SiSb system

A. Csik[a,b,*], G. Erdélyi[a], G.A. Langer[a], L. Daróczi[a], D.L. Beke[a], J. Nyéki[a], Z. Erdélyi[a]

[a]*Department of Solid State Physics, University of Debrecen, P.O. Box 2, Debrecen H-4010, Hungary*

[b]*On leave from the Institute of Nuclear Research, Hungarian Academy of Sciences (ATOMKI), P.O. Box 51, Debrecen H-4001, Hungary*

**Abstract**

Thermal annealing of $Si/Si_{1-x}Sb_x/Si$ amorphous thin film tri-layer samples (x=18 and 24 at%Sb) under 100 bar Ar pressure results in an interesting pattern formation. In pictures, taken by means of cross-sectional transmission electron microscopy (TEM), stripe-shaped contrast, with three maxima, parallel with the interfaces can be seen. Secondary neutral mass spectrometer (SNMS) measurements revealed that the regions with different contrasts correspond to Sb-rich and Sb-depleted regions. Furthermore, the Sb concentration peaks in the Sb-rich regions, especially at longer annealing times, are different: the peak developed at the Si/SiSb interface closer to the free surface decays faster than that of the inner one closer to the substrate. The pattern formation is interpreted by segregation-initiated spinodal-like decomposition, while the difference of the Sb concentration peaks is explained by the resultant Sb transport to and evaporation from the free surface. The possible role of formation of nanocrystalline grains, in the explanation of the fast transport under pressure as compared to vacuum is also discussed.

[*] Corresponding author. *Tel./Fax*: +36-52-316073
*E-mail address*: csik@atomki.hu (A. Csik)

**Introduction**

Recently much attention has been paid to study amorphous semiconductor materials because of their electronic and optical properties [1,2,3]. Besides the amorphous Si, the amorphous Si-based semiconductor alloys (for example Si-X, X=Ge, Al, Sn) have also been studied, because attractive properties can be achieved by means of alloying of the above elements. Some of them (Al, Sb, Sn) induce amorphous-crystalline transformation (metal-induced crystallization (MIC)), reducing considerably the crystallization temperature of the amorphous Si [4,5]. This phenomenon offers an economical way for producing polycrystalline Si films. The MIC was mainly studied in Si-Al system [5]. The effect of Sb on the crystallization of the amorphous Si was investigated by Hentzell et al. [4,7], but the details of the kinetics of Sb segregation and diffusion have not been revealed yet.

The Si-Sb phase diagram and our recent experimental results on Sb segregation in amorphous Si system indicate that the mixing enthalpy in this system is positive [6,7]. The phase separation process in such amorphous systems may occur either by nucleation and growth, or by spinodal decomposition.

In our previous work [8] we investigated the transformations in amorphous $Si_{1-x}Sb_x$ mono-, $Si/Si_{1-x}Sb_x/Si$ tri- and $Si/Si_{1-x}Sb_x$ multilayers of different compositions (x=13-26 at%) under different hydrostatic pressures and vacuum by cross-sectional TEM. Annealing the mono- and multilayer samples in Ar atmosphere, the Sb containing layers became crystalline at 883 K, while the pure Si layers remained amorphous under those conditions. The Sb content and the applied hydrostatic pressure enhanced the crystallization and decomposition processes. The most interesting result was that under pressure in the Si/SiSb/Si trilayers (and only in this type of samples and only under hydrostatic pressures) the SiSb layer underwent a spinodal-like decomposition, resulting in three Sb-rich stripes parallel to the interfaces. No such stripes could be detected when the samples of the same geometry and composition were annealed in vacuum. The phenomenon was interpreted by an amorphous-nanocrystalline transformation (which enhanced the effective diffusivity) and by segregation initiated decomposition. Similar pattern formation was found experimentally during spinodal decomposition of polymer films [9] or recently in some metal-silicate systems [10]. Simulations also indicated that layered, striped structure can develop in the near surface/interface region, termed as surface-directed spinodal decomposition [11,12].

In this study – in contrast to [8], where only qualitative indications on the composition wave formation were presented - we investigate the distribution of the components by means of SNMS technique. In order to detect the possible crystallization processes we carry out X-ray and electron diffraction measurements as well.

**Experimental**

Amorphous Si/SiSb/Si tri-layer were deposited by alternate sputtering of pure elements [Si (99.999 at%) and Sb (99.99 at%)] in a dc magnetron system onto Si (100) substrates held at room temperature. The base pressure was $3 \times 10^{-7}$ mbar and the Ar operating pressure was $5 \times 10^{-3}$ mbar. The amorphous $Si/Si_{1-x}Sb_x/Si$ trilayer-films (x=18 and 24 at %) with thickness d(Si)=20 nm and d(SiSb)=40 nm were deposited on Si wafers. Annealing treatments were performed at 723-823 K in vacuum (p<$10^{-7}$ mbar) and in Ar atmosphere at 100 bar. SNMS and XRD measurements were performed in every 30 minutes, interrupting the heat treatment. The temperature was measured by a NiCr-Ni thermocouple attached to the sample holder and controlled within ±1%.

Microstructural characterization of the films was carried out after annealing, using a standard XRD spectrometer with Cu-cathode and JEOL 2000 transmission electron microscope equipped with an energy-dispersive X-ray spectrometer (Oxford Link-Isis EDS). The concentration profile was measured by a SNMS type SPECS INA-X.

**Results and discussion**

In our previous paper [8] the pattern formation was detected annealing at 883 K, t=120 minutes, Ar pressure: 100 bar, sample geometry: Si 50nm/Si$_{74\ at\%}$Sb$_{26\ at\%}$ 50 nm/Si 50 nm, (Fig. 1). In our recent experiments we used samples of the following geometry and composition: Si 20nm/Si$_{82\ at\%}$Sb$_{18\ at\%}$ 40 nm/Si 20 nm. Having annealed the samples under 100 bar Ar pressure, at temperatures 723, 773, 823 K for 2 hours, TEM pictures revealed that pattern formation takes place at temperatures of 773 and 823 K. By means of electron diffraction no crystallization could be detected. This result suggests that the decomposition of the alloy can start already in amorphous state, since the crystallization temperature of this alloy is about 873 K [13]. Investigating the samples showing pattern by SNMS, the stripes could be identified as Sb-rich regions (Fig. 2).

In order to see the effect of concentration, we also prepared samples with 24 at% Sb in the central layer. Heat treatments were carried out at 723 K. X-ray diffraction measurements could not detect any crystallization after 1170 minutes long heat treatment, i.e. the films remained X-ray amorphous. A for the pattern formation, the SNMS investigations revealed that composition maxima developed in that sample as well. Concentration peaks were observed at the original Si/SiSb and SiSb/Si interfaces, furthermore, it was also observed that a segregated Sb-rich thin layer was formed at the topmost surface (Fig. 3). It was also detected that the Sb peaks are asymmetrical; the peak on the substrate side is generally higher than the one closer to the topmost surface (Fig.2). Fig. 3 also illustrates that with increasing time this asymmetry becomes even more expressed, indicating that there is a Sb transport to the free surface. The driving force of the diffusion is the Sb gradient established very likely by the evaporation of Sb from the free surface. Finally, this out-diffusion of the Sb leads to the decay of the Sb peaks developed at the original interfaces. It is interesting that the enrichment of the Sb at the topmost surface changes with the annealing time, see Fig.3.

Our results confirmed that in both type of samples (18 and 24 at% Sb), the first detectable process is the segregation of Sb at the interfaces including the segregation at the free surface as well. Considering that in our system phase separation tendencies are expected, spinodal decomposition may result similar Sb concentration profiles as observed experimentally. In Fig. 4 simulation results on spinodal decomposition in a thin slab are displayed [12]. The initial segregation may initiate crystallization near the interfaces, the formation of nano-size crystalline grains increases the segregated amount of Sb. Though we could not detect crystallization, such process can not be ruled out, since the amorphous–nanocrystalline transformation can hardly be revealed by X-ray diffraction. Our experimental finding that the decomposition was never observed in vacuum, could qualitatively be interpreted supposing that the pressure may enhance the crystallization [8]. This supposition is supported by the density data of the amorphous and crystalline phases: the crystalline phase has lower density than the amorphous one [8].

The number of the composition peaks depends on the thickness of the slab, the segregation behavior, solubility limits, etc. In order to find experimental evidence to support the spinodal-like decomposition we should investigate how the concentration profiles are affected by the thickness and the composition of the SiSb layer. These experimental investigations and theoretical simulations are in progress.

**Conclusion**

We have used TEM, SNMS, X-ray and electron diffraction to show the evolution of the microstructure of Si/SiSb/Si trilayers during thermal annealing in vacuum and under 100 bar Ar pressure. In Si/SiSb/Si trilayers (only under pressure) the SiSb layer underwent a spinodal-like decomposition resulting in three stripes parallel to the interfaces. The SNMS investigations revealed that asymmetrical composition maxima developed at the original Si/SiSb and SiSb/Si interfaces and that a segregated Sb-rich thin layer is formed at the topmost surface. Our results confirmed that the first step of the process is the segregation of Sb at the interfaces and later a spinodal decomposition results in the formation of composition peaks. The asymmetry of the Sb peaks was explained by the resultant Sb transport to and evaporation from the free surface.

**Acknowledgements**

This work was supported by OTKA Grant No. D-048594, T-043464, T-038125, F-043372 and "Bolyai János" scholarship (Z. Erdélyi).

**Figure capture**

Figure 1. Cross-sectional TEM image of Si50 nm/ Si$_{74\%}$Sb$_{26\%}$50 nm/Si50nm trilayers annealed at 883 K, for 2 h in 100 bar Ar [8].

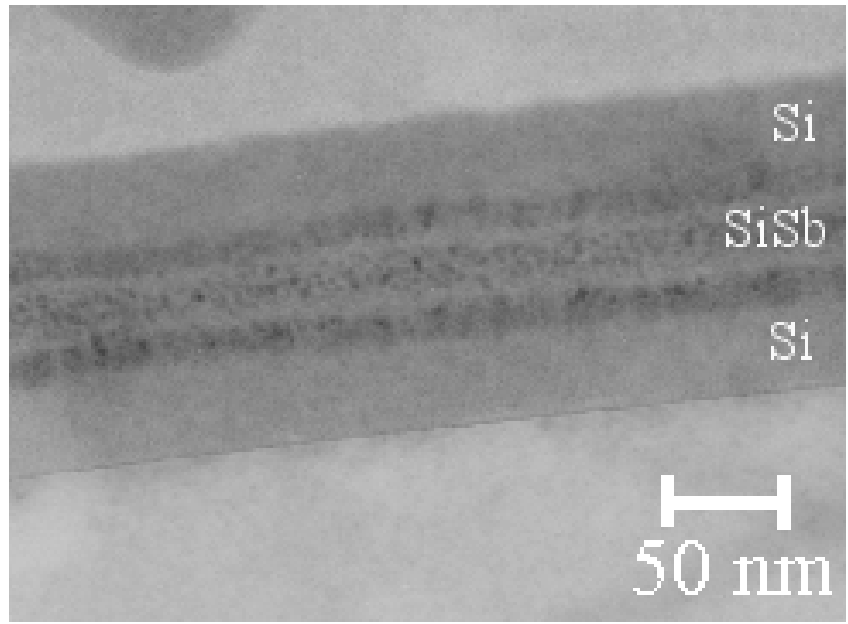

Figure 2. The distribution of Sb in Si20nm/Si$_{82at\%}$Sb$_{18at\%}$40nm/Si20nm tri-layer measured by SNMS.

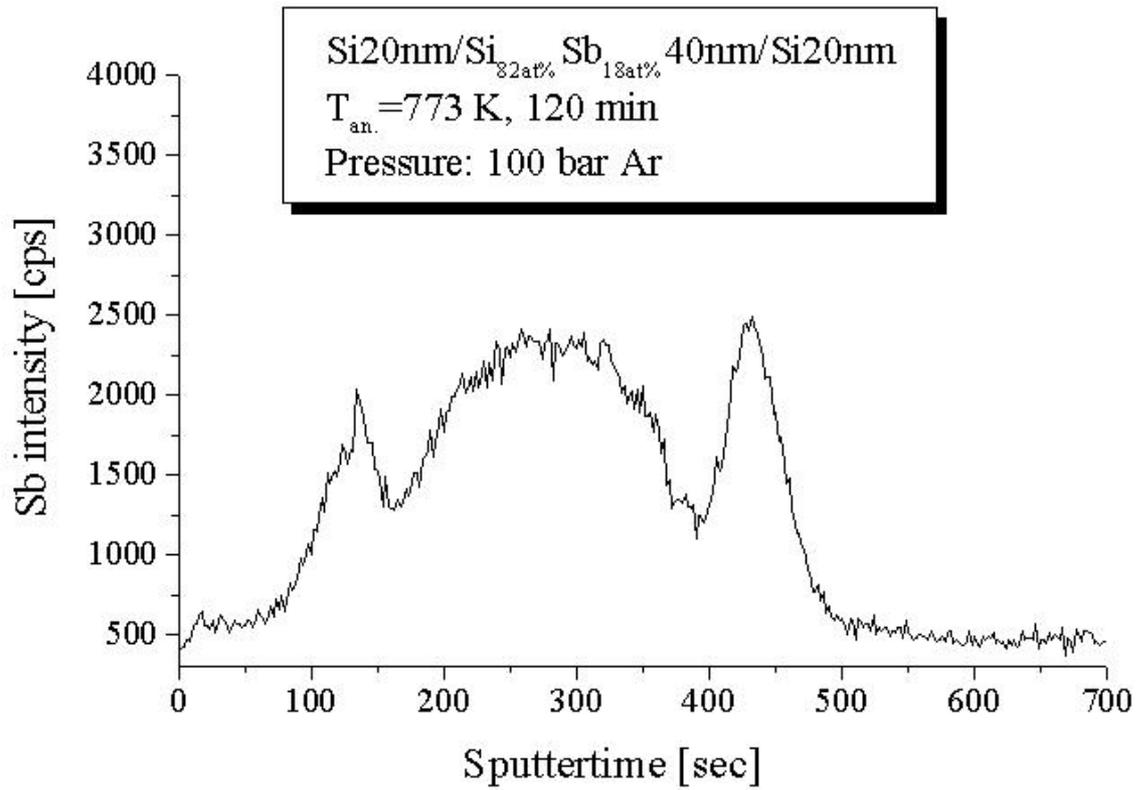

Figure 3. SNMS concentration profiles of Si20nm/Si$_{76at\%}$Sb$_{24at\%}$40nm/Si20nm trilayer measured at different annealing time. The enrichment of the Sb at the topmost surface (at zero sputtertime) changes with the annealing time.

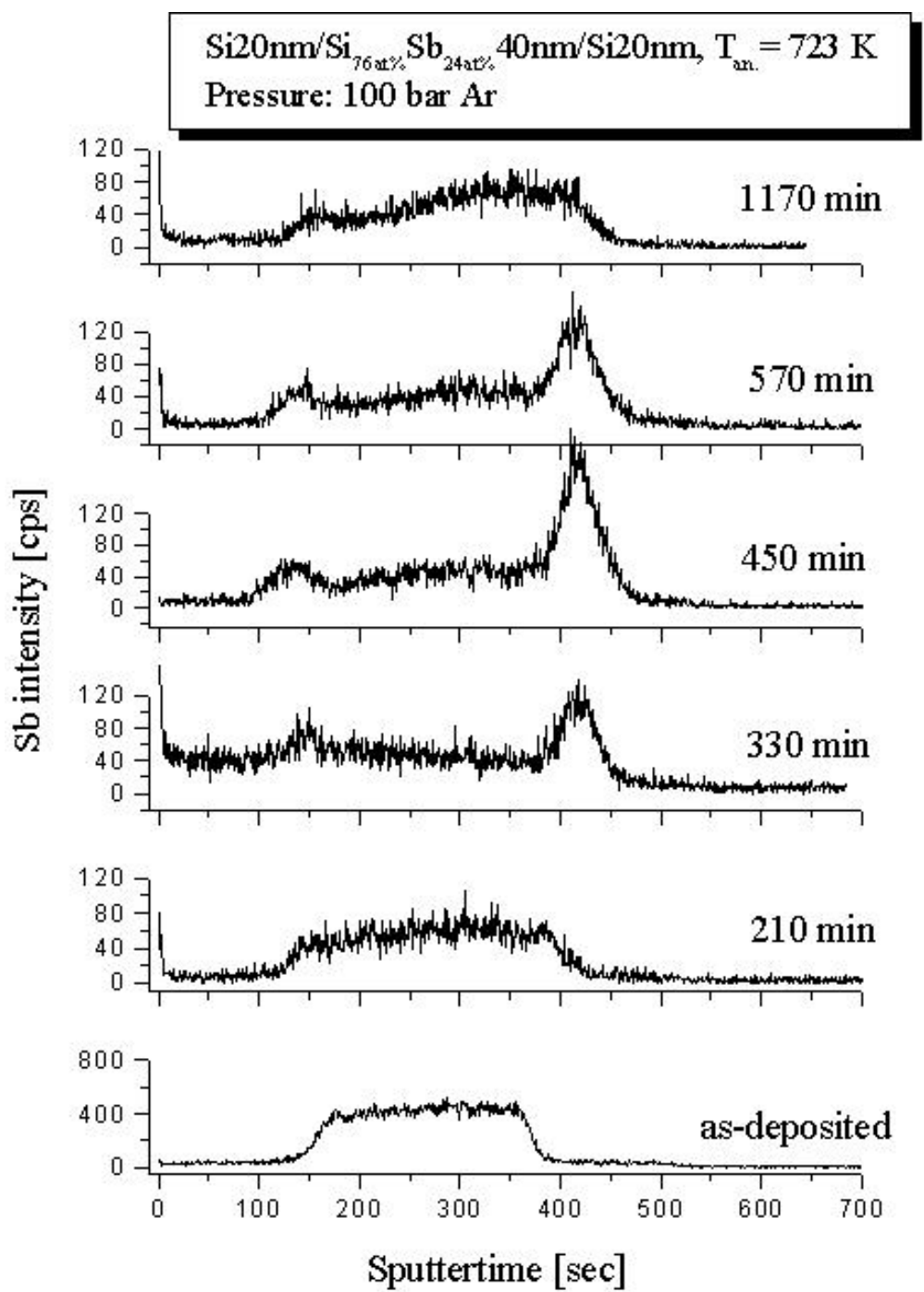

Figure 4. Simulated time evaluation of composition profile in the Cu/V system for two different film thicknesses (increasing numbers at the curves indicated the time sequence) [12].

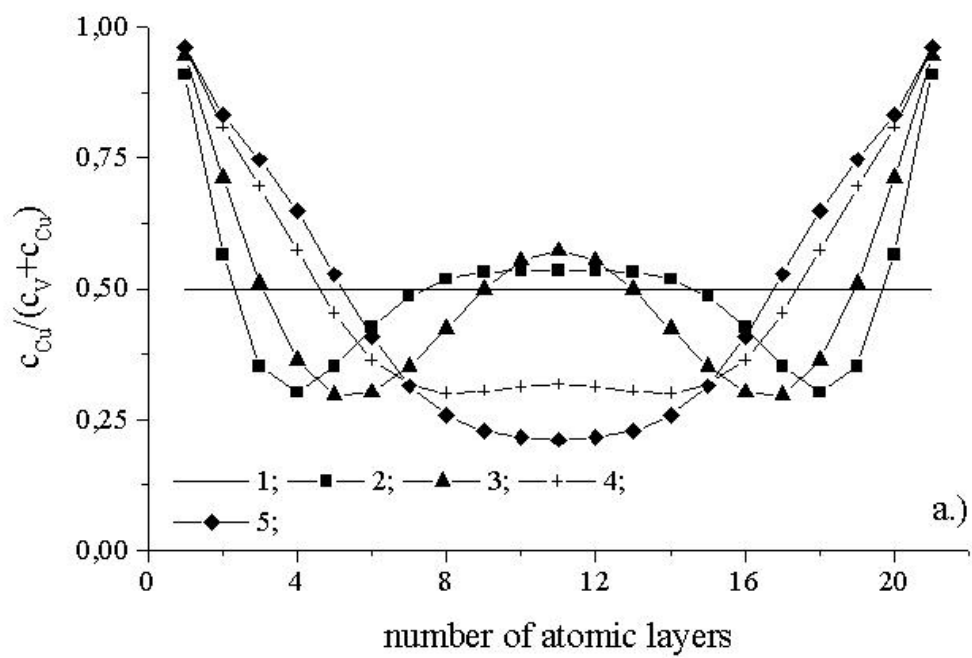

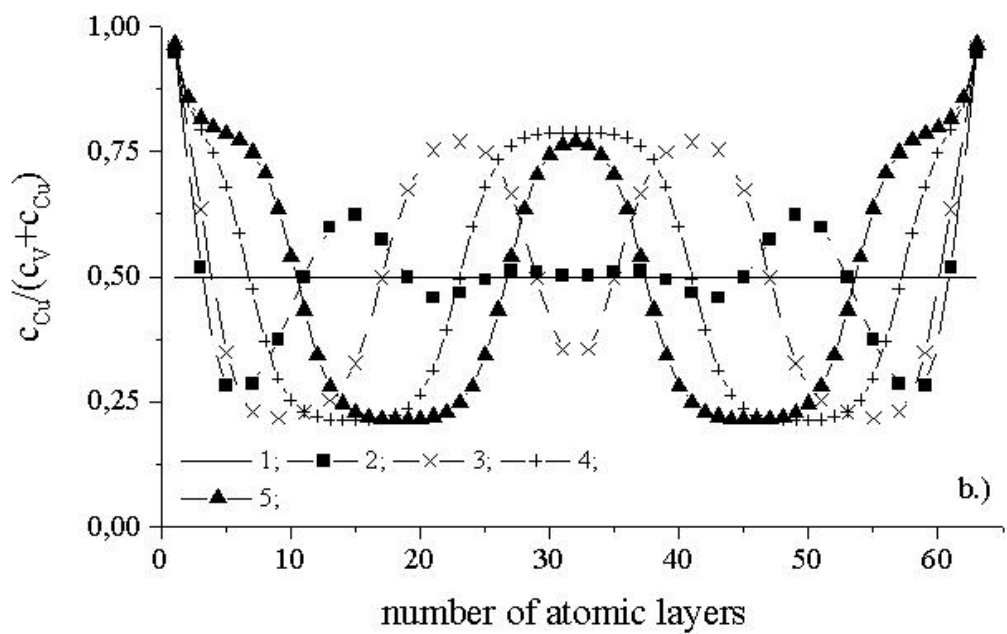